# A Study on the Expanding Universe Based on a Model of the Time Variation of its Matter Content in the Framework of Brans-Dicke Theory

Sudipto Roy[1], Dibyajyoti Laha[2], Argho Aranya Sangma[2], Indrani Pal[2]

[1]Department of Physics, St. Xavier's College, Kolkata

[2] Batch of 2014-16, M.Sc. in Physics, St. Xavier's College, Kolkata

30 Mother Teresa Sarani (Park Street), Kolkata – 700016, West Bengal, India.

[1]email:   roy.sudipto1@gmail.com

**Abstract:** A theoretical model of cosmic expansion has been formulated on an assumption of inter-conversion of matter and dark energy, in the framework of Brans-Dicke theory. An empirical scale factor has been used, which generates a signature flip of the deceleration parameter with time. To account for the non-conservation of matter, a function of time f(t) is incorporated into the equation representing the density of matter. Its value at any instant of time is proportional to the matter content of the universe. The functional form of f(t) has been determined from the field equations by using an empirical scalar field parameter expressed in terms of the scale factor. It is found to decrease with time almost monotonically, implying a conversion of matter into dark energy. Using this function f(t), the time variation of the density of matter has been determined and also the expressions regarding the proportions of matter and dark energy of the universe have been formulated. Time variation of gravitational constant, its fractional rate of change and the Brans-Dicke dimensionless parameter has been analyzed. The dependence of Brans-Dicke parameter upon the scalar field has been determined. The present study enables us to correlate the change of matter content with the change of deceleration parameter and gravitational constant without using any specific mechanism of interaction between matter and scalar field.

## 1.     Introduction

Recent astrophysical observations have strongly confirmed that the universe is expanding with acceleration (Bennet *et al.*, 2003; Riess *et al.*, 1998; Perlmutter *et al.*, 1999). Several studies have shown that nearly 70% of the constituents of the universe has a large negative pressure and is referred to as dark energy (DE), and it is generally believed to be the sole cause of accelerated expansion of the universe. It has not yet been possible to determine its true nature. A widely known parameter of the general theory of relativity, the cosmological constant ($\lambda$), is one of the most suitable candidates acting as the source for this repulsive gravitational effect and it fits the observational data reasonably well, although it has its own limitations (Sahni and Starobinsky, 2000). A large number of possible candidates for this dark energy component has already been proposed and their behaviour have been studied extensively (Sahni and Starobinsky, 2000; Padmanabhan, 2003). It is worth mentioning that this accelerated expansion is a very recent phenomenon and it follows a phase of decelerated expansion. This is important for the successful nucleosynthesis and also for the structure formation of the universe. As per observational findings, beyond a certain value of the redshift (z) (i.e. $z > 1.5$), the universe surely had a decelerated phase of expansion (Riess *et al.*, 2001). So, the dark energy component should have evolved in such a way that its effect on the dynamics of the universe is dominant only during later stages of the matter dominated era. A recent study, based on an analysis of supernova data, by Padmanabhan and Roy Choudhury (2003), has shown that the deceleration parameter ($q$) of the universe has certainly changed its sign from positive to negative, indicating a change of phase from deceleration to acceleration.

In addition to the models using the cosmological constant ($\lambda$), a large number of other models of dark energy have already appeared in the literature with their own characteristic features (Copeland *et al.*, 2006; Martin, 2008). All these theoretical formulations have been found to generate the cosmic acceleration very effectively. Out of these candidates, one of the most popular candidates is a scalar field with a positive potential which can generate an effective negative pressure if the potential term dominates over the kinetic term. This scalar field is often called the *quintessence scalar field*. One finds a large number of quintessence potentials in scientific literature and their behaviours have been studied extensively. To get detailed information in this regard, one may go through a study by V Sahni (2004) on this field. The origins of models of most of the quintessence potentials do not have a proper physical background or explanation.

To determine properly the role played by such entities, one should consider that the two components of matter, namely, the cold dark matter (CDM) and dark energy or the Q-matter are interacting among themselves and, consequently, there would be some transfer of energy from one field to another. A number of models have been proposed where a transfer of energy takes place from the component of dark matter to the component of dark energy (Zimdahl, 2012; Reddy and Kumar, 2013), so that during the late time of evolution, the dark energy predominates over the ordinary matter and causes acceleration of the universe. However, most of these models are found to be based on interactions that have been chosen arbitrarily, without being supported by any physical theory. To represent the role of dark energy there has been a prolonged search for a cosmologically viable model formulated on the basis of an interaction of matter with the scalar field.

Due to the ambiguities and difficulties caused by the arbitrariness in the formulation of a particular Q-field, non-minimally coupled scalar field theories have been shown to be much more capable of carrying out the possible role of an agent responsible for the late time acceleration. This is simply due to the presence of the scalar field in the purview of the theory and does not have to be introduced separately. The Brans-Dicke (BD) theory is considered to be the most natural choice as the scalar-tensor generalization of general relativity (GR), due to its simplicity and a possible reduction to GR in some limit. For this reason, Brans–Dicke theory or its modified versions have been shown to generate the present cosmic acceleration (Banerjee and Pavon, 2001a; Brunier *et al.*, 2005). It has also been shown that BD theory can potentially generate sufficient acceleration in the matter dominated era even without any help from an exotic Q - field (Banerjee and Pavon, 2001b). But one actually needs a theory which can account for a transition from a state of deceleration to acceleration. The dark energy and dark matter components are considered to be non-interacting in most of the models and are allowed to evolve independently. Since these two components are of unknown nature, the interaction between them is expected to provide a relatively generalized framework for study. Zimdahl and Pavon (2004) have shown that the interaction between dark energy and dark matter can be very useful in solving the coincidence problem. Following this idea, one may consider an interaction or inter-conversion of energy between the Brans - Dicke scalar field which is a geometrical field and the dark matter. The possibility of an inter-conversion of energy between the matter content of the universe and the non-minimally coupled scalar field had been predicted earlier by Amendola (1999).

In the framework of Brans-Dicke theory, it has been found that in most of the models the accelerated expansion of the universe requires a very low value of $\omega$, typically of the order of unity (Das *et al.*, 2014). In one of such studies it was shown that, if the BD scalar field interacts with the dark matter, a generalized BD theory can perhaps serve the purpose of driving acceleration even with a high value of $\omega$ (Banerjee and Das, 2006). In all these studies, either Brans-Dicke theory is modified to meet the present requirement or, one chooses a quintessence scalar field to generate the required acceleration. In a recent study by Barrow and Clifton (2006) and also in the reference (Banerjee and Das, 2006), it was shown that no additional potential was required to cause the signature flip (from positive to negative) of the deceleration parameter. They used an interaction between the BD scalar field and the dark matter to explain the observational findings in this regard.

In the present study, a generalized form of Brans-Dicke theory has been used. It was proposed by Bergman (1968) and expressed in a more useful form by Nordtvedt (1970). Here, the BD dimensionless parameter $\omega$ is regarded as a function of the BD scalar field $\varphi$ instead of being treated as a constant. This study is not based on any particular theoretical model regarding the mechanism of conversion of matter into dark energy. Here we propose a model which only takes into account the fact that the matter content of the universe is not conserved, keeping the possibilities open for an inter-conversion between matter and some other entity, probably the dark energy which is regarded as the sole agent that causes the accelerated expansion of universe. To describe the time dependence of the density of matter ($\rho$), we have incorporated a function $f(t)$ in its expression, in order to account for the non-conservation of matter. It is quite evident from this equation that if we consider $f(t) = 1$ at all time, the expression of $\rho$ represents conservation of the matter content of the universe. In Brans-Dicke field equations, we have used an empirical expression of scale factor ($a$) and scalar field parameter ($\varphi$) to obtain the expressions of $f(t)$. According to its definition, $f(t)$ is proportional to the matter content of the universe at any time $t$. We have explored the behaviour this function to get an estimate of the time evolution of matter and dark energy of the universe, assuming the dark energy to have been produced entirely from matter. We have studied the time variation of mater density and the gravitational constant ($G$) and also the time dependence of the proportions of matter and dark energy of the universe. This study enables us to correlate the change of matter content with the change of deceleration parameter and gravitational constant. As the rate of dark energy production decreases, the rate of change of gravitational constant and deceleration parameter are also found to decrease.

## 2. Theoretical Model

The field equations in the generalized Brans-Dicke theory, for a spatially flat Robertson-Walker space-time, are given by (Banerjee and Ganguly, 2009),

$$3\left(\frac{\dot{a}}{a}\right)^2 = \frac{\rho}{\varphi} + \frac{\omega(\varphi)}{2}\left(\frac{\dot{\varphi}}{\varphi}\right)^2 - 3\frac{\dot{a}}{a}\frac{\dot{\varphi}}{\varphi}, \quad (1)$$

$$2\frac{\ddot{a}}{a} + \left(\frac{\dot{a}}{a}\right)^2 = -\frac{\omega(\varphi)}{2}\left(\frac{\dot{\varphi}}{\varphi}\right)^2 - 2\frac{\dot{a}}{a}\frac{\dot{\varphi}}{\varphi} - \frac{\ddot{\varphi}}{\varphi}. \quad (2)$$

Combining (1) and (2) one gets,

$$2\frac{\ddot{a}}{a} + 4\left(\frac{\dot{a}}{a}\right)^2 = \frac{\rho}{\varphi} - 5\frac{\dot{a}}{a}\frac{\dot{\varphi}}{\varphi} - \frac{\ddot{\varphi}}{\varphi}. \quad (3)$$

From equation (1), the expression of $\omega(\varphi)$ is obtained as,

$$\omega(\varphi) = 2\left[3\left(\frac{\dot{a}}{a}\right)^2 - \frac{\rho}{\varphi} + 3\frac{\dot{a}}{a}\frac{\dot{\varphi}}{\varphi}\right]\left(\frac{\dot{\varphi}}{\varphi}\right)^{-2} \quad (4)$$

There are many theoretical studies where the content of matter (dark + baryonic) of the universe has been assumed to be conserved (Banerjee and Ganguly, 2009). This conservation of matter is mathematically expressed as,

$$\rho a^3 = \rho_0 a_0^3 = \rho_0 \quad (\text{taking } a_0 = 1) \quad (5)$$

There are some studies on Brans-Dicke theory of cosmology where one takes into account an interaction between matter and the scalar field. A possibility of an inter-conversion between dark energy and matter (both dark and baryonic matter) is taken into consideration in these studies. Keeping in mind this possibility, we propose the following relation for the density of matter ($\rho$).

$$\rho a^3 = f(t)\rho_0 a_0^3 = f(t)\rho_0 \quad (\text{taking } a_0 = 1) \quad (6)$$

In the present study we have not considered any theoretical model to explain or analyse the mechanism of interaction between matter and the scalar field. We have only considered a simple fact that the right hand side of equation (5) cannot be independent of time when one takes into account non-conservation of matter due to its generation from dark energy or its transformation into dark energy. We propose to introduce a function of time $f(t)$ in equation

(5) to get a new relation represented by equation (6). This function $f(t) = \frac{\rho a^3}{\rho_0 a_0^3}$, at any instant of time $t$ is the ratio of matter content of the universe at the time $t$ to the matter content at the present instant ($t = t_0$). Thus $f(t)$ can be regarded as proportional to the total content of matter (dark+baryonic) $M(t)$ of the universe at the instant of time $t$. We have denoted this ratio by $R_1$ where $R_1 \equiv f(t) = M(t)/M(t_0)$. We have defined a second ratio $R_2 = \frac{1}{f}\frac{df}{dt} = \frac{1}{M}\frac{dM}{dt}$ which represents fractional change of matter per unit time. If, at any instant, $R_2$ is negative, it indicates a loss of matter or a change of matter into some other form due to its interaction with the scalar field. We have also defined a third ratio $R_3 = f - 1 = \frac{M(t)-M(t_0)}{M(t_0)}$ indicating a fractional change of matter content from its value at the present time.

One may assume that the process of conversion of matter into dark energy started in the past at the time of $t = \gamma t_0$ where $\gamma < 1$. Hence, $M(\gamma t_0) = M(t_0)R_1(\gamma t_0)$ is the total matter content of the universe, at $t = \gamma t_0$, when no dark energy existed. Thus $M(\gamma t_0)$ is the total content of matter and dark energy at all time. Assuming matter to be the only source of dark energy, the proportion of dark energy in the universe at any time $t$ is given by the following ratio ($R_4$).

$$R_4 = \frac{M(\gamma t_0) - M(t)}{M(\gamma t_0)} = \frac{f(\gamma t_0) - f(t)}{f(\gamma t_0)} \qquad \gamma < 1 \tag{7}$$

Thus ($R_4 \times 100$) is the percentage of dark energy present in the universe. Nearly 70% of the total matter-energy of the universe is dark energy at the present time (Das and Mamon, 2014). For a proper choice of $\gamma$ and $k$ (to be defined later), we must have $R_4(t_0) \times 100 = 70$ approximately.

The proportion of matter (dark + baryonic) in the universe is therefore given by,

$$R_5 = 1 - R_4 = 1 - \frac{M(\gamma t_0) - M(t)}{M(\gamma t_0)} = \frac{M(t)}{M(\gamma t_0)} = \frac{f(t)}{f(\gamma t_0)} \tag{8}$$

Thus ($R_5 \times 100$) is the percentage of matter (ordinary+dark) present in the universe.
The purpose of the present study is to determine a functional form of $f(t)$ to explore the time dependence of the ratios $R_1, R_2, R_3, R_4$ and $R_5$.

Using these parameters, the density of dark energy can be expressed as,

$$\rho_d = \frac{R_3}{R_4}\rho \tag{9}$$

To formulate the expression of $f(t)$ we have used the following relation which is based on equation (6).

$$f(t) = a^3 \frac{\rho}{\rho_0} \tag{10}$$

Here, the density of matter ($\rho$) can be obtained from equation (3). For this purpose one needs to choose some suitable functional form of the Brans-Dicke scalar field $\varphi$. In the present study we have chosen an empirical forms of $\varphi$, following some recent studies in this regard (Banerjee and Ganguly, 2009; Roy et al., 2013). The proposed ansatz for $\varphi$ is expressed as,

$$\varphi = \varphi_0 \left(\frac{a}{a_0}\right)^k = \varphi_0 a^k \tag{11}$$

Here $k$ is a constant which determines the rapidity with which the parameter $\varphi \left(\equiv \frac{1}{G}\right)$ changes with time.

Combining equation (11) with equation (3), one gets the following expression of density of matter of the universe ($\rho$).

$$\rho = \varphi H^2 [k^2 + (4-q)k + (4-2q)] \tag{12}$$

Using equation (12), $\rho_0$ can be written as,

$$\rho_0 = \varphi_0 H_0^{\,2} [k^2 + (4-q_0)k + (4-2q_0)] \tag{13}$$

Substituting from equations (12) and (13) into equation (10) we get,

$$f(t) = a^3 \frac{\varphi H^2 [k^2 + (4-q)k + (4-2q)]}{\varphi_0 H_0^{\,2} [k^2 + (4-q_0)k + (4-2q_0)]} \tag{14}$$

In our derivation of the equations (12) we have used the standard expressions of Hubble parameter ($H$) and deceleration parameter ($q$), which are, $H = \dot{a}/a$ and $q = -\ddot{a}a/\dot{a}^2$ respectively. In the expression of $f(t)$, in the equation (14), the parameters $\varphi$, $H$ and $q$ are all functions of time. Their time evolution depends upon the time variation of the scale factor from which they are calculated. To calculate $f(t)$ using equation (14), we have used an empirical scale factor. This scale factor has been chosen in order to satisfy a recent observation regarding the deceleration parameter $q(\equiv -\ddot{a}a/\dot{a}^2)$. According to this observation the universe had a state of decelerated expansion before the present phase of acceleration began (Banerjee and Ganguly, 2009; Banerjee and Das, 2006; Das and Mamon, 2014). Thus, the deceleration parameter had a positive value before reaching the present stage of negative values. The functional form of our chosen scale factor is such that the deceleration parameter, calculated from it, shows a change of sign as a function of time. This scale factor is,

$$a = a_0 \, Exp\left[-\alpha t_0^\beta\right] Exp[\alpha t^\beta] \tag{15}$$

Here $\alpha, \beta$ are constants. Here $\beta > 0$ to make sure that the scale factor increases with time. The scalar field parameter ($\varphi$), Hubble parameter ($H$) and deceleration parameter ($q$), based on this scale factor are given by,

$$\varphi = \varphi_0 \left(\frac{a}{a_0}\right)^k = \varphi_0 Exp\left[-k\alpha t_0^\beta\right] Exp[k\alpha t^\beta] \tag{16}$$

$$H = \dot{a}/a = \alpha \beta t^{(\beta-1)} \tag{17}$$

$$q = -\ddot{a}a/\dot{a}^2 = -1 + \frac{1-\beta}{\alpha\beta} t^{-\beta} \tag{18}$$

Here we find that, for $0 < \beta < 1$ and $\alpha > 0$ we have,

$q > 0$ for $t = 0$ and $q \to -1$ as $t \to \infty$

It clearly means that the chosen scale factor generates an expression of deceleration parameters which changes sign from positive to negative as time goes on. The values of constant parameters $(\alpha, \beta)$ have been determined from the following conditions.

Condition 1: $H = H_0$ at $t = t_0$ \hfill (19)

Condition 2: $q = q_0$ at $t = t_0$ \hfill (20)

Combining the equations (19) and (20) with the equations (17) and (18) respectively, one obtains
$\alpha = 6.39 \times 10^{-10}$ and $\beta = 5.37 \times 10^{-01}$

To determine the value of $f(t)$ from the equations (14), one must use the expressions of (16), (17) and (18) and also the values of the constants $\alpha$ and $\beta$.

The function $f(t)$ is defined by the relation $\rho a^3 = f(t) \rho_0 a_0^3$. According to this relation, the value of $f(t)$ is always positive and, $f(t) = 1$ at $t = t_0$ (taking $a_0 = 1$). The functional form $f(t)$ in equation (14) ensures that $f(t) = 1$ at $t = t_0$. The values of $k$ for which $f(t)$ is positive over the entire range of study (say, from $t = 0.5t_0$ to $t = 1.5t_0$) is given below.

$k < (k_-)_{min}$ or $k > (k_+)_{max}$ over the entire range of study. Here,

$$(k_-)_{min} = (q-2)_{min} \quad \text{over the range from } t = 0.5t_0 \text{ to } t = 1.5t_0 \tag{21}$$

$$(k_+)_{max} = (q-2)_{max} \quad \text{over the range from } t = 0.5t_0 \text{ to } t = 1.5t_0 \tag{22}$$

For our range of study, i.e. from $t = 0.5t_0$ to $t = 1.5t_0$, we find,

$(k_-)_{min} = -2.64$ and $(k_+)_{max} = -2.33$

Thus we have a lower and an upper range of permissible values for $k$ which are $k < (k_-)_{min}$ or $k > (k_+)_{max}$ respectively. The upper range, $k > (k_+)_{max}$, includes both positive and negative values of $k$ and the lower range, $k < (k_-)_{min}$, has only negative values. According to equation (11), the parameter $\varphi$ is a decreasing function of time for negative values of $k$, causing the gravitational constant $(G = \frac{1}{\varphi})$ to increase with time. Therefore we find that the upper range of $k$ allows $G$ to be both increasing and decreasing function of time, although the lower range of $k$ causes $G$ to be an increasing function of time. To choose between these two ranges of $k$, we have to determine the values of $\omega_0$ at different values of $k$ and compare them with those obtained from other studies. Using equation (4) we can write the following expression (eqn. 23) regarding $\omega$ for this model.

$$\omega = \frac{2}{k^2}\left[3(1+k) - \frac{\rho}{\phi H^2}\right] \quad (23)$$

Using equations (23) we get the following expression of $\omega_0$ (the value of $\omega$ at the present epoch).

$$\omega_0 = \frac{2}{k^2}\left[3(1+k) - \frac{\rho_0}{\phi_0 H_0^2}\right] \quad (24)$$

According to several studies on Brans-Dicke theory $\omega_0$ has a negative value close to $(-1)$ (Das and Mamon, 2014).

To have $\omega_0 < 0$, the condition to be satisfied by $k$ is given by,

$$k < \frac{\rho_0}{3\,\phi_0 H_0^2} - 1 \quad \text{or,} \quad k < -0.9884 \quad (25)$$

For the entire lower range of $k$ values and for a part of its upper range, the above condition is satisfied.

The gravitational constant ($G$), which is reciprocal of the Brans-Dicke scalar field parameter ($\varphi$) is given by,

$$G = \frac{1}{\varphi} = \frac{(a/a_0)^{-k}}{\varphi_0} = \frac{1}{\varphi_0} Exp[k\alpha t_0^\beta]\, Exp[-k\alpha t^\beta] \quad (26)$$

An experimentally measurable parameter $\frac{\dot{G}}{G}$ is given by,

$$\frac{\dot{G}}{G} = \frac{1}{G}\frac{dG}{dt} = -k\frac{\dot{a}}{a} = -kH \quad (27)$$

Using equation (27) we get,

$$\left(\frac{\dot{G}}{G}\right)_{t=t_0} = -kH_0 \quad (28)$$

The value of $k$ should be so chosen that $\left(\frac{\dot{G}}{G}\right)_{t=t_0} < 4 \times 10^{-10}\ Yr^{-1}$ (Weinberg, 1972).

## 3. Graphical analysis of theoretical findings

We have plotted $\omega_0$ as a function of $k$ in Figure 1. For the lower range of $k$ values, the values of $\omega_0$ are negative and also close to the values obtained from other studies (Das and Mamon, 2014). For the upper range of $k$, we have both positive and negative values of $\omega_0$, the positive values being totally contrary to the findings of other studies in this regard (Das and Mamon, 2014). This observation is in favour of using the lower ranges of $k$ values.

A plot of $(\dot{G}/G)_{t=t_0}$ as a function of $k$ is shown in Figure 2. Its positive value indicates that $G$ is presently increasing with time and negative value indicates its reverse. For the entire lower range of $k$ values it is positive. For the upper range of $k$ values it is found to be both positive and negative. There are experimental observations and theoretical models where $G$ has been shown to be increasing with time (Pradhan *et al.*, 2015).

It is evident from Figure 3 that the Brans-Dicke parameter $\omega$ is an increasing function of time, with only negative values, for the lower range of $k$ values. But its behavior is different for the upper range of $k$ values. For the positive values of the upper range of $k$, the values of $\omega$ are all positive which is inconsistent with the findings of other studies (Das and Mamon, 2014).

On the basis of these findings from the figures 1, 2 and 3, we have found it logical to chose values from the lower range of $k$ to determine the time dependence of $f(t)$ and other relevant parameters connected to it in the present study.

Figure 4 shows the variation of $R_1 (\equiv f)$ as a function of time for three different values of $k$ in its lower range. It shows that the matter content of the universe $[M(t) = f(t)M_0]$ decreases with time and the rate of its reduction is faster for more negative values of $k$.

The variation of $R_2$ and $R_3$ as functions of time for two different values of $k$, in its lower range, has been shown in Figure 5. It shows that $R_2$, the fractional rate of change of matter content is always negative, implying a transformation from matter to other energy forms. Its value decreases with time, indicating a slowing down of the transformation process as time goes on. As $k$ values become more negative, transformation rate becomes larger. The values of $R_3$ are positive in the past ($t < t_0$) and negative in future ($t > t_0$), as expected from the time variation of $R_1$ and $R_2$. It also changes faster for more negative values of $k$.

Figure 6 shows the plots of $R_4$ and $R_5$ as functions of time. The proportion of dark energy ($R_4$) in the universe is found to increase with time. Since it is assumed to be generated at the cost of matter (dark + baryonic), the proportion of matter in the universe ($R_5$) decreases with time. Thus, the sum of $R_4$ and $R_5$ is unity since the beginning of the conversion process at $t = \gamma t_0$.

Two of the three plots of figure 7 show the variation of matter density ($\rho$) of the universe as a function of time for two different values of $k$ in its lower range. These plots are based on equation (6). These curves show that the density decreases at a faster rate for more negative values of the parameter $k$. The third plot of this figure is based on equation (5) which shows the variation of matter density when the matter is assumed to remain conserved. For the green curve we have assumed that $f(t) = 1$ for all values of $t$. Plots corresponding to the non-conservation regime shows a faster fall than that for the conservation of matter. For the third plot ($f = 1$), the density of matter changes only due to the expansion of the universe. For two other plots, the density changes due to the expansion of the universe and also the conversion of matter into some other form of energy.

Figure 8 shows the variation of $q$ and $G/G_0$ as functions of $R_4$, the proportion of dark energy of the universe. As $R_4$ increases, the deceleration parameter becomes more negative, indicating greater acceleration of the cosmic expansion. The gravitational constant increases as the dark energy proportion of the universe increases. Therefore the dark energy may be assumed to have a role in causing accelerated expansion and also in increasing the gravitational constant.

All these plots show that as $k$ is made more and more negative, the rate of reduction of matter content (proportional to $R_1$) and its density ($\rho$) increases and the rate of change of gravitational constant and deceleration parameter increases. It indicates clearly that there exists a relation of the transformation of matter into dark energy with the gravitational constant and state of expansion of the universe characterized by the deceleration parameter and its time variation.

## 4. Conclusions

The main objective of the present study is to determine of the time dependence of the density of matter, both dark and ordinary matter, of the universe by incorporating a suitable combination of empirical expressions of the scale factor and the scalar field parameter in Brans-Dicke field equations for zero curvature space. The expression of the density of matter, used in the field equation, has been constructed on an assumption of inter-conversion between matter and dark energy. The density of matter has been taken to be dependent upon a function of time $f(t)$, which is proportional to the matter content of the universe at any instant of time. While incorporating this function into the expression of density, we have not assumed any form of this function. We have determined its form from the field equations. To improve this model one may choose a particular form of this function $f(t)$, which is chosen in this model to account for the non-conservation of matter-energy. It is found that this function, which is proportional to the matter content of the universe, decreases with time, clearly implying a direction of conversion, that is from the matter component to the component of dark energy of the universe. It is found through graphical analysis that, as the proportion of dark energy content of the universe increases with time, the deceleration parameter becomes more and more negative and the gravitational constant is found to increase with time. All these observations are pointing towards a possibility that there is a dependence of the gravitational constant and the deceleration parameter on the content of dark energy and the rate at which it changes. It has been shown in the present study that if the dark energy is assumed to be produced completely at the cost of matter (both dark and baryonic), its present proportion should depend on the time at which the process of conversion of matter into energy began in the universe. The present study shows that, by a proper tuning of parameters, it is possible to obtain values from the theoretical model, which are close to the experimental evidence regarding the present proportion of dark energy and matter. As part of a future extension of this study, this model can be modified by changing the empirical form of the Brans-Dicke scalar field parameter ($\varphi$) and also by selecting a better empirical form of $f(t)$ in terms of the scale factor and the scalar field. While choosing the functional form of $f(t)$ one must make sure that its value is unity at the present epoch, according to its defining relation $\rho a^3 = f(t)\, \rho_0 a_0^3$. One may study the time dependence of several such forms of $f(t)$ to determine the nature of conversion of matter into dark energy and the way it influences the process of cosmic expansion by changing the gravitational constant and deceleration parameter with time.

# FIGURES

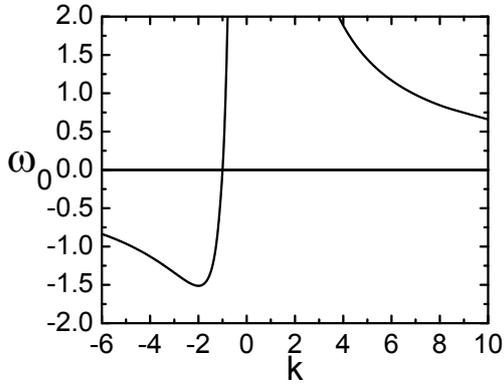

Figure 1: Plot of the Brans-Dicke parameter $\omega$ at $t = t_0$ as a function of $k$.

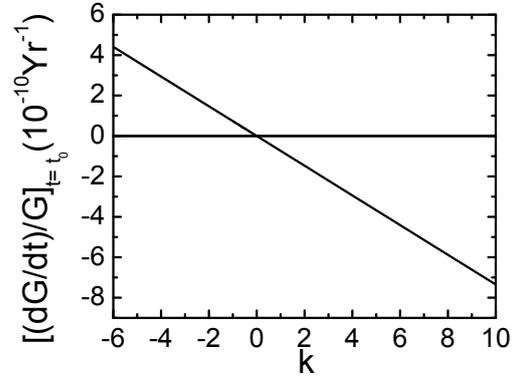

Figure 2: Plot of $(\dot{G}/G)_{t=t_0}$ ( in $10^{-10} Yr^{-1}$) as a function of $k$.

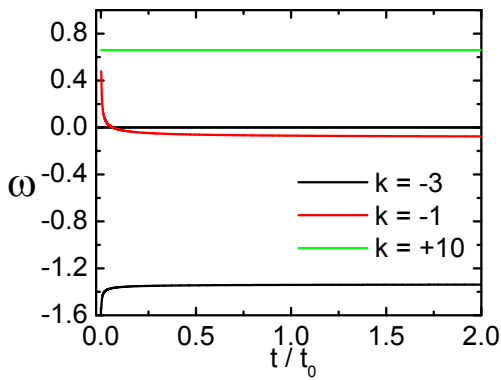

Figure 3: Plot of $\omega$ as a function of time for upper (*red, green*) and lower (*black*) ranges of $k$ values.

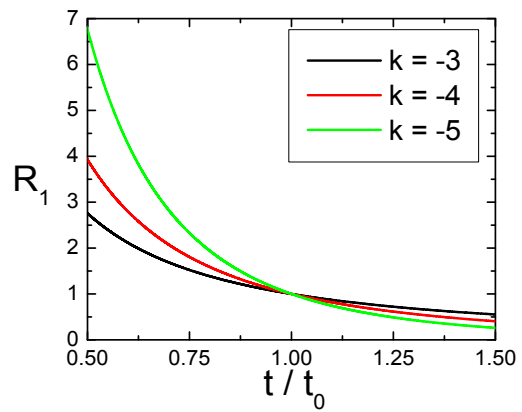

Figure 4: Plot of $R_1 (\equiv f)$ as a function of time for three different values of $k$ in its lower range.

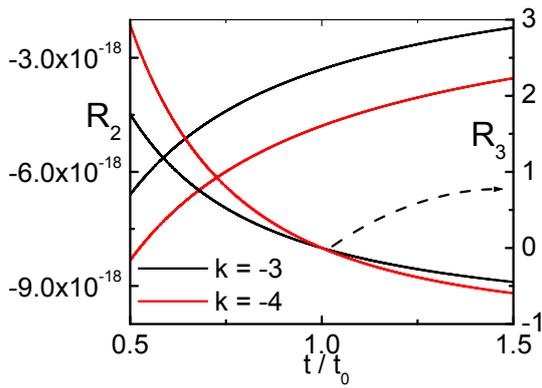

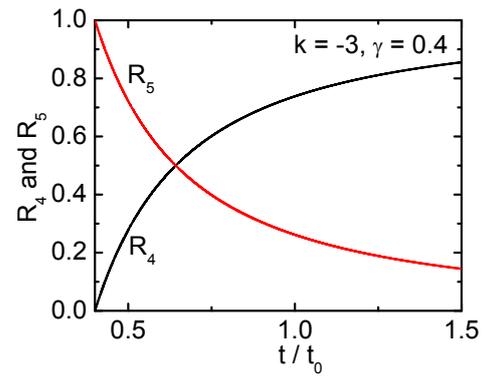

Figure 5: Plot of $R_2$ (*increasing*) and $R_3$ (*decreasing*) as functions of time for $k = -3$ (*black*) and $k = -4$ (*red*). $R_2$ and $R_3$ are along left and right vertical axes respectively.

Figure 6: Plot of $R_4$ (*black*) and $R_5$ (*red*) as functions of time for $k = -3$ and $\gamma = 0.4$. Matter to energy conversion began at $t = \gamma t_0$.

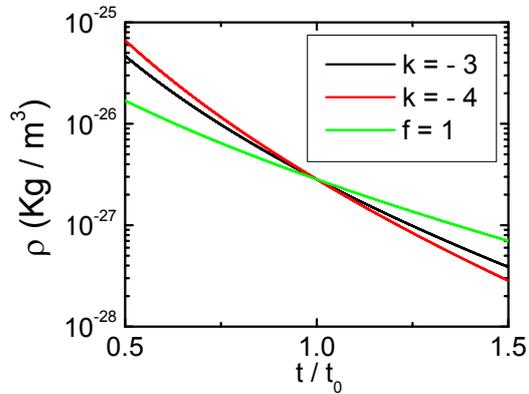

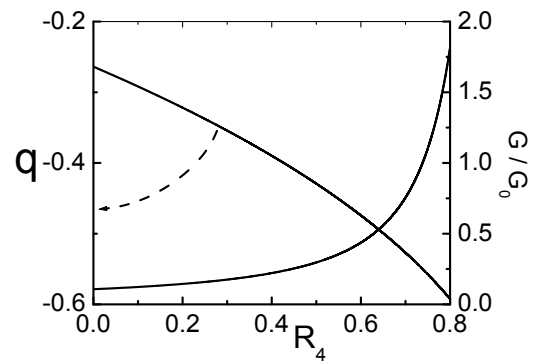

Figure 7: Plot of $\rho$ as a function of time for conservation ($f = 1$) and non-conservation ($k = -3,4$) of matter.

Figure 8: Plot of $q$ and $G/G_0$ as functions of $R_4$ along the left and right vertical axes respectively for $k = -3$ and $\gamma = 0.4$.